# Magnetic flux penetration and AC loss in a composite superconducting wire with ferromagnetic parts

F. Gömöry, M. Vojenčiak, E. Pardo, J. Šouc

Institute of Electrical Engineering, Centre of Excellence CENG, Slovak Academy of Sciences, Dúbravská cesta 9, 842 39 Bratislava, Slovakia

**Abstract:** The current distribution and the AC loss in a composite superconducting tape containing a layer from magnetic material is calculated and compared with experiments, showing a very good agreement. The situations of an alternating uniform applied field or a transport current are studied. The newly developed numerical model is an approximation to the critical state model, adapted for the applicability to commercial finite elements codes that solve the vector potential. Substantial feature of this procedure is that it can be carried out in the case when the critical current density in superconductor depends on the magnetic field and the magnetic layer material is non-linear. Additionally, the hysteresis loss in the magnetic material is estimated, based on its measured magnetization loops. Measurements on Bi-2223 multifilamentary tapes covered on edges by nickel confirmed our predictions, showing a substantial ac loss reduction in both the investigated regimes.

**Short title**: AC loss in superconductor-ferromagnetic composite wire

1. Introduction

The electromagnetic performance of a superconducting wire in AC applications is determined by two main features: the maximum current transported without resistance, and the rate at which the electromagnetic energy is converted into heat. The quantity used to indicate the first characteristic is the critical current, $I_c$. The second property is commonly characterized by the AC loss dependence on transport current amplitude, $I_a$ (transport or self-field loss) and on the amplitude of the applied magnetic field, $B_{max}$ (magnetization loss). These dependences are a good measure of the tape quality. It has been found experimentally [1] that the performance of a commercial Bi-2223/Ag tape in these two regards can be improved with the help of a ferromagnetic covering of the tape edges. This kind of sample represents a more general case of a wire containing both superconducting (SC) and ferromagnetic (FM) layers. It is therefore of broader interest to investigate to what extent we can theoretically predict the behaviour of such SC-FM composite and optimize its design.

The observed increase of critical current in low background magnetic fields has been explained by numerical simulations. This situation corresponds to the case when all the cross-section of the wire is filled with the maximum current density that the material could carry without resistance, i.e. the critical current density, $j_c$ . Detailed knowledge of the dependence of $j_c$ on both the magnitude and orientation [2] of the local magnetic field is essential when reliable predictions should be achieved [3]. With these data, the state-of-the-art finite element codes are able to resolve the saturation of superconducting part of the composite with electrical current. In such static calculation the superconductor is characterized by a nonlinear $\boldsymbol{j_c}(\boldsymbol{B})$ dependence and the ferromagnetic cover by its nonlinear permeability $\boldsymbol{\mu}(\boldsymbol{B})$. We have successfully used the empirical expressions

$$\boldsymbol{j_c(x,y)} = \frac{\boldsymbol{j_{c0}}}{\left(\boldsymbol{1} + \frac{\sqrt{\boldsymbol{k^2 B_{||}^2(x,y) + B_{\perp}^2\,(x,y)}}}{\boldsymbol{B_0}}\right)^{\boldsymbol{\beta}}} \tag{1}$$

and

$$\boldsymbol{\mu(x,y)} = \boldsymbol{\mu_0}\left[\boldsymbol{1} + \frac{\boldsymbol{\mu_{r,max}}}{\boldsymbol{1} + \left(\frac{|\boldsymbol{B(x,y)}|^2}{\boldsymbol{B_c^2}}\right)^{\boldsymbol{\alpha}}}\right] \tag{2}$$

where $\boldsymbol{\mu_0} = \mathbf{4\pi} \times \mathbf{10^{-7}\ H/m}$ to resolve the distributions of current density and magnetic field in the *xy*-section of the wire with the longitudinal dimension along the *z*-coordinate of the Cartesian system - see Figure 1. Parallel and perpendicular components of the magnetic field, $B_{||}$ and $B_{\perp}$, respectively, are taken with regard to the orientation of the longer dimension of the superconducting core. In the case of high temperature superconductor Bi-2223, this is also the orientation with respect to the *a-b* planes in superconducting grains that is commonly parallel to the wide face of the superconducting tape. Note that for the complete description of SC properties the establishing of four parameters $j_{c0}$,, $k$, $B_0$ and $\beta$ is necessary and the expression (2) for FM permeability contains three parameters $\mu_{r,max}$, $B_c$ and $\alpha$.

In difference to a rather reliable prediction of the critical current, the calculation of AC loss requires to know how the process of filling the superconductor with (critical) current density develops during the cycle of AC current or AC magnetic field. There have been powerful numerical methods invented to resolve this problem for superconductors of various shapes [4, 5]. In these calculations the superconductor is divided in filamentary elements, and the process of flux penetration is equivalent to filling them with current. The matrix of mutual inductances between elements is calculated in the preparatory part, and further on it governs the evolution of current distribution. Adding a FM layer with constant permeability requires to modify the calculation of mutual inductances. Using the finite element procedure for this purpose, it was found that the FM layer could significantly modify the way in which magnetic flux penetrates the wire [6] and the loss in superconducting part can be higher or lower, depending on the exact geometry of SC-FM composite [7]. This is also the finding from analytical calculations [8].

Main drawback of assuming a constant FM permeability is the impossibility to estimate the loss contribution because of FM magnetic hysteresis. In the attempts published until now, this problem was treated separately from the calculation of the current and field distribution in superconductor [9].

From the practical point of view, the calculation of AC loss based on a commercial numerical code is the procedure that at most could benefit from rapid development of hardware and software. Its availability to cope with non-linear FM properties is already a standard. A very interesting approach has been recently done by A. Campbell [10] that resolves the evolution of critical state in hard superconductor in the way suitable for finite element codes. Following that work but starting from an alternative physical justification we have developed a procedure that is particularly suitable for numerical calculations of magnetic flux penetration into SC-FM composite with non-linear properties of both superconducting and magnetic part. The AC loss in superconductor can be evaluated, and plausible estimation of the loss due to FM hysteresis performed as well. Summing these two components we obtained the total loss that should be compared with experimental data. As will be shown later in this paper, experiments confirmed the validity of our calculation method.

## 2. Critical state formulation in 2D using the vector potential of magnetic field

In this section, we formulate the theoretical model from which the numerical calculations are developed, taking as a starting point the critical state model. The reader not interested in the physics beyond the proposed procedure can go directly to the Section 3.

Here we show how our method works in two commonly used arrangements: In the pure transport case, the actual value of the electrical current $I$ carried by the wire is the imposed parameter driving the change of current and field distribution inside the wire. When a transversal AC magnetic field is applied to an insulated wire, the change of its actual value, $B_a$, causes the re-arrangement of the field and current distributions.

The geometry considered here is two-dimensional (2D): an infinite wire along the $z$-axis, with the distribution of current and field in any $x$-$y$ section identical – see Figure 1. In this geometry, the vector potential of magnetic field, $\vec{\boldsymbol{A}}$, is the chosen independent variable for electromagnetic calculation. Its curl is the source of magnetic flux density

$$\vec{\boldsymbol{B}} = \nabla \times \vec{\boldsymbol{A}} \tag{3}$$

and the equation to be resolved is

$$\boldsymbol{\nabla}\times\left(\frac{\mathbf{1}}{\boldsymbol{\mu}}\boldsymbol{\nabla}\times\vec{\boldsymbol{A}}\right)=\vec{\boldsymbol{j}} \tag{4}$$

The magnetic permeability, $\mu$, as well as the current density, $\vec{\boldsymbol{j}}$, can be functions of the position $\vec{\boldsymbol{r}}$ = (x,y) and of the magnitude of magnetic field.

The critical state model [12] basically states, that in a hard superconductor one can find the local current density with the magnitude that is either zero or equal to the critical current density. Zero current density appears in the part that never experienced any electrical field. The orientation of the current density is controlled by the most recent non-zero electrical field existing locally in the superconductor. The electrical field accompanying the change of the distribution of electrical current and magnetic field is, in general

$$\vec{\boldsymbol{E}}=-\frac{\boldsymbol{\partial}\vec{\boldsymbol{A}}}{\boldsymbol{\partial t}}-\boldsymbol{\nabla}\boldsymbol{\varphi}. \tag{5}$$

Equation (5) is valid for any gauge of the vector potential.

Let us now focus our consideration to superconductors following the critical state model. For simplicity, in this article we restrict to the situations when (i) there exist a zone where $\vec{\boldsymbol{E}}$ vanishes in all the cycle – "neutral zone" – and (ii) the magnetic flux penetrates monotonically from the surface when $I$ or $B_a$ increase monotonically. The first condition is always true except the case of transporting a current equal or larger than the critical current, a situation not significant for our AC loss calculation[1]. The second condition is fulfilled in many practical situations, including our solutions in Section 3.

In the case of superconductor infinitely extended along the *z*-axis, also the electrical field components in the *x*-*y* plane must vanish because of symmetry reasons. Thus $\nabla\varphi$ has only the *z*-component, in other words the value of $\varphi$ is the same in all the cross-section of the wire. Moreover, since $\vec{\boldsymbol{E}}$ is independent on *z*, $\nabla\varphi$ is uniform; it only depends on time. Applying then the equation (5) to the so called "neutral zone" with $\vec{\boldsymbol{E}} = \boldsymbol{0}$ one finds that the vector potential follows

$$\boldsymbol{\nabla\varphi} = -\frac{\partial\overrightarrow{\boldsymbol{A_c}}}{\partial\boldsymbol{t}}, \qquad (6)$$

where $\overrightarrow{\boldsymbol{A_c}}$ is the vector potential at the neutral zone. For our electromagnetic calculations, we use the calibrated vector potential

$$\vec{\boldsymbol{A}}' = \vec{\boldsymbol{A}} - \overrightarrow{\boldsymbol{A_c}} \qquad (7)$$

in further computations. Its properties are summarized as follows[2]:

---

[1] For a large applied magnetic field (and no transport current) the sample is fully penetrated by current but the condition (i) is still fulfilled because there is a boundary between $j=j_c$ and $j=-j_c$, where $j=0$ and, consequently, $E=0$ (figure 2).

[2] In the most general case, (8d) contains a uniform time independent (constant) term due to static electrical current charge distributions. In our case, we do not consider DC transport currents and this contribution to $E$ vanishes.

$$\vec{A}'_c = \mathbf{0} \tag{8a}$$

$$\nabla \times \vec{A}' = \nabla \times \vec{A} \tag{8b}$$

$$\vec{E} = -\frac{\partial \vec{A}'}{\partial t} \tag{8c}$$

In the following we will analyse the process of a change in transport current or applied field in terms of the vector potential $\vec{A}'$. Because the magnetic vector potential as well as the electric field and current density have only the *z*-components $A_z$ , $E_z$, and $j_z$, respectively, we can omit the vectorial notation of these quantities. Moreover, in the infinitely long geometry, the flux per unit length between two points in the *xy* plane is the vector potential difference. Since $A'$ in the neutral zone is zero, $A'(x,y,t)$ is the magnetic flux per unit length between the neutral zone and the point $(x,y)$.

## 2.1 Initial part

Let us now consider the first step in exciting the superconductor from the virgin state, i.e. from $j = 0$ everywhere. This means in transport case that the total current jumps as $\mathbf{0} \to \boldsymbol{I}_{ac,1}$, and in the magnetization case the applied field changes as $\mathbf{0} \to \boldsymbol{B}_{ac,1}$. The values of vector potential locally changed from the original distribution $\boldsymbol{A}'_{\mathbf{0}}(\boldsymbol{x},\boldsymbol{y})$ to the actual distribution $\boldsymbol{A'_1(x,y)}$. From the expression (8c) follows that the zero current density would remain in the parts of the superconductor where the vector potential has not changed, and vice versa:

$$\boldsymbol{j_1(x,y)} = \begin{cases} +\boldsymbol{j_c} & \boldsymbol{if\ A'_1(x,y) > A'_0(x,y)} \\ -\boldsymbol{j_c} & \boldsymbol{if\ A'_1(x,y) < A'_0(x,y)} \\ \boldsymbol{0} & \boldsymbol{if\ A'_1(x,y) = A'_0(x,y)} \end{cases} \tag{9}$$

When in the next step the transport current increases as $\boldsymbol{I_{ac,1}} \rightarrow \boldsymbol{I_{ac,2}}$, or the applied magnetic field grows as $\boldsymbol{B_{ac,1}} \rightarrow \boldsymbol{B_{ac,2}}$, the same principle of comparing the local values of vector potential controls the resulting distribution of current density. Next, we use the assumption that the magnetic flux penetrates monotonically in the superconductor for a monotonic increase of $I$ or $B_a$. Since $A'(x,y)$ is the magnetic flux per unit length between the neutral zone and $(x,y)$, this means that $A'$ increases monotonically in the initial stage. Therefore:

$$\boldsymbol{j_{ini}(x,y)} = \begin{cases} +\boldsymbol{j_c} & \boldsymbol{if\ A'(x,y) > A'_p(x,y)} \\ -\boldsymbol{j_c} & \boldsymbol{if\ A'(x,y) < A'_p(x,y)} \\ \boldsymbol{0} & \boldsymbol{if\ A'(x,y) = A'_p(x,y)} \end{cases} \tag{10}$$

Here, the local value of the (calibrated) vector potential achieved in the previous step of the change of transport current or applied field has been denoted as $\boldsymbol{A'_p}$ . The notion (10) could be simplified utilizing the signum function

$$\boldsymbol{j_{ini}(x,y) = j_c\ \mathrm{sign}\big(A'(x,y) - A'_p(x,y)\big)} \tag{11}$$

As already suggested by Campbell [10], for the purpose of numerical calculations it is favourable to replace the signum function exhibiting infinite derivatives at zero with a smoother equivalent. The form we have found perfectly plausible for our finite element code is the hyperbolic tangent, with the independent variable scaled by a factor that on one side should secure the convergence of calculations and on the other side should be compatible with the physical formulation of the problem – see Figure 2. Thus the equation we inserted in the finite element calculation to define the current density in superconductor is

$$\boldsymbol{j}_{ini}(\boldsymbol{x},\boldsymbol{y}) = \boldsymbol{j}_c \ \mathbf{tanh}\left(\frac{\boldsymbol{A'}_{\boldsymbol{p}}(\boldsymbol{x},\boldsymbol{y}) - \boldsymbol{A'}(\boldsymbol{x},\boldsymbol{y})}{\boldsymbol{A}_{\boldsymbol{n}}}\right) \tag{12}$$

where the choice of the scaling factor $A_n$ influences the thickness of the borders between parts filled with opposite current densities.

In the case of a monotonic increase considered up to now there is no difference in assuming one single step from the virgin state with $\boldsymbol{j} = \mathbf{0}$ everywhere to the maximum current $I_a$ (or magnetic field $B_{max}$ ) and the case when the maximum value is reached through a series of positive steps.

## 2.2 Decreasing part of the cycle

Now let us extend our considerations to the part of the AC cycle with decreasing transport current or applied field. The difference with respect to the initial part described by the expression (10) is that a non-zero value of current density can be found also in the places with no actual change in the vector potential when it was induced there in one of the previous steps. That means we should put $j_p$, the value of $j$ achieved in the previous step, in place of zero in the third row of (10), leaving the first two rows unchanged. In this way we arrive to the central equation for the current density in superconductor, $j_s$,

$$\boldsymbol{j_s(x,y)} = \begin{cases} +\boldsymbol{j_c} & \boldsymbol{if\ A'(x,y) > A'_p(x,y)} \\ -\boldsymbol{j_c} & \boldsymbol{if\ A'(x,y) < A'_p(x,y)} \\ \boldsymbol{j_p(x,y)} & \boldsymbol{if\ A'(x,y) = A'_p(x,y)} \end{cases} \tag{13}$$

that is valid for any part of the AC. In this way we have found the right-hand side of the expression (4) that now can be used to resolve the distributions of $A(x,y)$ and $j(x,y)$ in self consistent way. Of course the used numerical procedure must allow to recall the distribution calculated in the previous step in the expressions controlling the actual distribution. Through this comparison the history of current distribution in superconductor is involved in very natural way. If $A'(x,y)$ decreases monotonically in one half cycle, $j_p$ in equation (13) can be left as $j$ at the $I$ or $B_a$ peak and the current distribution for the desired external parameter can be found in one step.

### 2.3 Boundary conditions

Essential ingredient of the finite element calculation is the proper choice of the boundary condition. We set them on the cylinder concentric with the superconducting tape, with a radius $R$ much larger than the tape width. This cylinder is the boundary box in the finite elements calculations in Section 3.

For the transport situation, we establish the boundary condition in the following way. In the Coulomb's gauge (defined as $\boldsymbol{\nabla\vec{A}' = 0}$), the vector potential created by the currents flowing in the wire, $I$, in a distance $R$ much larger than the dimensions defining the cross-section is

$$\boldsymbol{A}(\boldsymbol{R},\boldsymbol{t}) = -\frac{\boldsymbol{\mu_0 I_w(t)}}{\mathbf{2\pi}}\mathbf{ln}(\boldsymbol{R}) \tag{15}$$

Therefore, from (8b) we find that

$$\boldsymbol{A}'(\boldsymbol{R},\boldsymbol{t}) = -\frac{\boldsymbol{\mu_0 I_w(t)}}{\mathbf{2\pi}}\mathbf{ln}(\boldsymbol{R}) - \boldsymbol{A_c}(\boldsymbol{t}) \tag{16}$$

where $A_c$ is the vector potential in the kernel defined in the Coulomb's gauge. An important consequence of (16) is that the vector potential for any point on the cylinder, that conforms the boundary box, is the same. Since for our gauge conditions in (8), $A'(R,t)$ is the flux per unit length between the neutral zone and the boundary, $\Phi_R$, the boundary condition (16) becomes

$$\boldsymbol{A}'(\boldsymbol{R},\boldsymbol{t}) = -\boldsymbol{\Phi_R}(\boldsymbol{t}) \tag{17}$$

For a uniformly applied magnetic field, with $I = 0$, the term with a logarithm in (16) vanishes and $A_c$ for the Coulomb's gauge is also zero due to the mirror symmetry of our problem. In addition, the vector potential created by the uniform magnetic field $B_a$ applied in the $y$ direction is $-xB_a$. For this case, the vector potential from the applied field vanishes at the centre and it already follows our gauge conditions of (8). Therefore, the boundary condition is

$$\boldsymbol{A}'(\boldsymbol{R},\boldsymbol{t}) = -x\boldsymbol{B}_{\boldsymbol{a}}(\boldsymbol{t}) \quad (18)$$

in this case.

## 3. Procedure of finite element calculation

There are various ways to use the proposed approach for the numerical calculations of hysteresis in type II superconductors. Here we describe the procedure that was found reasonably fast and reliable in the case of AC loss calculation in a commercial Bi-2223 tape with elliptic filamentary zone, covered on edges by nickel. Considered geometry is shown in Figure 1. The wire is placed concentrically in the calculation box that is the infinite cylinder with radius $R$. We used the commercial finite element code [13] that was already found able to cope with nonlinearities of current density and magnetic permeability described by the formulas (1) and (2), respectively.

### 3.1 Applied magnetic field

In the case when the wire is exposed to a magnetic field we use the boundary condition (18). To determine the AC loss, a series of distributions should be calculated for finite number of applied magnetic fields corresponding to cyclic change of magnetic field.

We found it practical to perform the calculation in the following series of steps: First, the change of applied field from zero to the maximum value of the applied field, $B_a$, is considered in one single step. The choice of vector potential (18) suggests to assume $\boldsymbol{A'_p}(\boldsymbol{x}, \boldsymbol{y}) = \mathbf{0}$ as well as $\boldsymbol{j'_p}(\boldsymbol{x}, \boldsymbol{y}) = \mathbf{0}$ in this first calculation step when applying the general equation (12)

$$\boldsymbol{j_{s,ini}}(\boldsymbol{x}, \boldsymbol{y}) = \boldsymbol{j_c} \ \mathbf{tanh}\left(\frac{-\boldsymbol{A'}(\boldsymbol{x}, \boldsymbol{y})}{\boldsymbol{A_n}}\right) \tag{19}$$

In Figure 3 are shown four distributions calculated in this way as an example. The direction of current density in the right half is positive (because the vector potential of applied field is negative there according to the expression (18)) except the zones with no current where also the vector potential is zero. The value of the scaling factor used in the calculations was $A_n = 10^{-9}$ Vs/m. The dependence of $j_c(x,y)$ is given by the expression (1), with the parameters established from experimental data determined for the critical current in dependence on magnetic field of various orientations [14]. The permeability $\mu$ in the FM cover is given by the interpolation expression (2) while it is equal to the permeability of vacuum, $\boldsymbol{\mu_0}$, in both the superconductor and the free space. The parameters of the dependence (2) are derived from the magnetization measurement of a sample from the ferromagnetic material [15]. After finishing the first calculation, the distribution of vector potential at the field amplitude i.e. at $\boldsymbol{B_a} = \boldsymbol{B_{max}}$, is stored as $\boldsymbol{A'_{max}}(\boldsymbol{x}, \boldsymbol{y})$ . It is used for the subsequent calculation of current and field distributions at the values of applied magnetic field decreasing from $B_{max}$ down to $-B_{max}$.

Purely hysteretic nature of the magnetization process is reflected by omitting the time variable in the equations from (9) to (13). Particular choice of magnetic field values used to determine the distributions during the AC cycle therefore does not influence the AC loss evaluation. We found it convenient to use the linear ramp $\boldsymbol{B_{a,i}} = \boldsymbol{B_{max}}\left(\mathbf{1} - \frac{\mathbf{2}\boldsymbol{i}}{\boldsymbol{N}}\right)$ with $i = 1..N$ and $N$ being the total number of applied fields , $\boldsymbol{B_{a,i}}$, for which the calculation has been performed. There is no background magnetic field applied on top of the AC one, and the downwards half of the AC cycle is representative enough to supply the data for the whole cycle. We found it sufficient to calculate for 20 descending values of the field, i.e. $N = 20$, leading to the hysteresis curve containing 40 points in the whole AC cycle.

The advantage of using one half of the AC cycle in which the change of the applied field is monotonic is that one can use for the current density in superconductor the functional dependence quite similar to the one used for the initial part of the cycle:

$$\boldsymbol{j}_{s,down}(\boldsymbol{x},\boldsymbol{y}) = \boldsymbol{j}_c \tanh\left(\frac{\boldsymbol{A'}_{max}(\boldsymbol{x},\boldsymbol{y}) - \boldsymbol{A'}(\boldsymbol{x},\boldsymbol{y})}{\boldsymbol{A}_n}\right) \tag{20}$$

In this expression, the distribution $\boldsymbol{A'}_{max}(\boldsymbol{x},\boldsymbol{y})$ calculated at the AC field amplitude $B_{max}$ is used as the input in calculation of the distribution at $B_a < B_{max}$. Suitable choice of the calibration for the vector potential through the boundary condition (18) allowed the reduction of the general expression (10) to this form. The series of distributions calculated in this way for $B_{max}$ = 12 mT and four values of descending field $B_a$ = (6, 0, -6, -12) mT is shown in Figure 4. For each distribution we determined the magnetic moment per unit length of the wire, $m$, and the magnetization $\boldsymbol{M} = \boldsymbol{m}/\boldsymbol{S}_{SC}$ where $S_{SC}$ is the superconductor's cross-section. Because the applied magnetic field is in the $y$-direction, the magnetization contributing to the AC loss is

$$\boldsymbol{M}(\boldsymbol{B}_{ac},\boldsymbol{B}_a) = \frac{\mathbf{1}}{\boldsymbol{S}_{SC}}\int_{S_{SC}} -\boldsymbol{x}\,\boldsymbol{j}(\boldsymbol{x},\boldsymbol{y})\mathrm{d}\boldsymbol{S}_{SC} \tag{21}$$

In this expression, we used the result of Brandt and Indenbom [11] who have shown that replacing the ½ factor in the original formula for the magnetic moment is equivalent to accounting for the contribution due to currents closing the current loops at the ends of the wire.

The plot of magnetization data in dependence on the applied field gives the magnetization loops – see Figure 5. Evaluating the loop area allows to compute the volume density of AC loss per AC cycle – in J/m$^3$ - due to hysteresis of supercurrent distribution as

$$\boldsymbol{Q}_{\boldsymbol{mSC}}(\boldsymbol{B}_{\boldsymbol{a}}) = \oint \boldsymbol{M}(\boldsymbol{B}_{\boldsymbol{ac}}, \boldsymbol{B}_{\boldsymbol{a}})\mathbf{d}\boldsymbol{B}_{\boldsymbol{ac}} \tag{22}$$

In the case of a ferromagnetic cover without hysteresis of magnetization, this is the total AC loss of the composite wire. Later in this paper we show how the contribution of the hysteretic magnetization loop of the ferromagnetic material can be included in the total loss estimation.

### 3.2 Transport current

In the calculation of loss in the case of the wire transporting AC current the basic expressions (19) and (20) remain the same, but the boundary condition is (17). This means we impose the magnetic flux between the neutral zone and the boundary the procedure finds the current that fulfils this condition. However, it is usually intended to set a certain desired current in the superconductor. The relation between the flux and the current is not straightforward in the considered geometry of circular boundary and elliptic superconducting core. Generally, larger value of vector potential at the boundary would require larger transport current in the wire according to the relation (16). We have found practical to set a transport current approximate to the desired one $I_{des}$ by defining the vector potential at the boundary of the calculation box using

$$A'(R) = -\frac{\mu_0 I_{des}}{4\pi}\left(2\ ln\left(\frac{R}{r_{eq}}\right) + 1\right) \tag{23}$$

This expression corresponds to the flux between the centre of a round wire of radius $r_{\mathrm{eq}}$ and the cylinder with radius $R$ if the wire carries a transport current $I_{\mathrm{des}}$ uniformly distributed in its cross-section. In our case of the wire with elliptic cross-section defined by the semi-axes $a_{SC}$ and $b_{SC}$, respectively, we have used $r_{eq} = (a_{SC} + b_{SC})/2$ . Once the finite elements procedure finds the solution with the boundary condition (23), the actual transport current $I$ is found by simply integrating $j$ in the tape cross-section, resulting in a slightly different value than $I_{des}$.

Similarly as in the magnetization case treated above, we first calculate the distributions at initial increase of the current using the formula (19). In Figure 6 are illustrated the results of such calculation for the same composite wire as shown in the magnetization regime in Figure 3. Current density is of single (positive) polarity because the superconductor is filled with increasing total transport current. Any such distribution can be used as the starting point for calculation of distributions at the current decrease from $I_a$ to - $I_a$ through the set of intermediate values $I$. Now the expression (20) is to be used for the definition of current density, with $A'_{max}(x, y)$ stored from the calculation at the AC current amplitude $I_a$. Examples of the current distributions calculated in this way for transport currents decreasing from 50 A down are shown in Figure 7.

Note that for all these distributions except the case of the critical current shown down in Figure 6, there is a visible portion of superconductor free from current. According to the expression (8a) the vector potential in this "neutral zone" is equal to zero. For each transport current distribution, the significant quantity to be evaluated is the magnetic flux between the neutral zone and the line in certain distance from the wire centre. Analytical calculations for the wire without ferromagnetic cover state that the minimum distance suitable for the determination of magnetic flux relevant for the transport AC loss evaluation [16] is about 3 times the major ellipse of the superconducting core $a_{SC}$. We denote this quantity – per unit length of the wire - $\Phi_{3a}$. Its evaluation is straightforward in 2D distribution, because it is directly given by the difference of the vector potentials

$$\boldsymbol{\Phi_{3a}} = \boldsymbol{A'(3a_{SC}, 0) - A'_c = A'(3a_{SC}, 0)} \qquad (24)$$

The last simplification is allowed by the fact that in our calibration the vector potential in the neutral zone is zero. In Figure 8 left side are the hysteresis loops of $\Phi_{3a}(I_{ac})$ shown for four of the AC current amplitudes. The opening of the loops is not much manifested, because a large portion of magnetic flux is just proportional to the current – so called inductive or reactive flux prevails. The shape of the loop can be better inspected when the compensated flux $\boldsymbol{\Phi_c = \Phi_{3a} - L_1 I_{ac}}$ is plotted instead of $\Phi_{3a}$. This is shown in the right hand side of Figure 8. We chose the quantity $L_1$ (with dimension H/m) in such the way that the loops at small current amplitudes become horizontal [17].

The area of the loop gives the value of transport AC loss per cycle and unit length of the wire

$$\boldsymbol{Q_{tSC}} = \oint \boldsymbol{\Phi_{3a}} \mathbf{d}\boldsymbol{I_{ac}} \tag{25}$$

Remember this is the loss due to hysteresis of current distribution in superconductor, that in our case is influenced by the presence of the ferromagnetic cover on edges. We did not assume a hysteresis magnetization loop for the FM material in our calculation procedure, just its nonlinearity as described by the expression (2). Comparison with experiments on real samples would require the knowledge of loss in ferromagnetic material. We discuss this issue in the next Section.

## 4. Comparison with experiments

In this paper we present the numerical and experimental data obtained for the commercial Bi-2223/Ag tape with 50 A self-field critical current at 77 K [15]. The 30 μm layer of nickel was deposited on the edges of the tape by galvanic process, leaving the central part of the tape uncovered in the width of 1.9 mm. Optical micrograph of the tape is shown in Figure 9 together with the geometrical representation used in the calculations. The filamentary zone was replaced by an effective superconducting core with elliptic cross-section, defined by the major half-axis $a_{SC}$ = 1.504 mm and the minor half-axis $b_{SC}$ = 0.11 mm, respectively. Declination of the Ni-cover by the angle $\delta$ = 4.3º was also taken into account in the calculations.

In order to establish the parameters of the $j_c(B)$ dependence, the procedure detailed in Reference [14] was followed. It is based on critical current data, $I_c(B)$, taken on the original non-covered tape in the range of applied magnetic fields from 0 to 0.15 T at several inclinations from 0 (parallel field) up to 90 degrees (perpendicular field). For the whole set of experimental conditions, numerical calculations were performed assuming all the central elliptic core is filled with critical current density defined by the expression (1). The optimal set of parameters in this expression is found at the condition of minimal discrepancy between experimental data and numerical results. In this way, the parameters characterizing the properties of the elliptic superconducting core have been determined as $j_{c0} = 1.34\times10^8$ A/m$^2$, $B_0 = 0.008$ T, $k = 0.1$ and $\beta = 0.58$.

To establish the properties of ferromagnetic Ni-layer deposited on Ag matrix, a 10 μm layer of Ni was deposited on a 1 mm thick silver plate 10 mm long by the same procedure as used in covering the tape edges. Magnetization of this sample in parallel magnetic field was measured at 77 K in SQUID magnetometer [15]. From these data, the dependence of permeability on magnetic field shown in Figure 10 (left side) was extracted. In the same plot is given the approximation in form of the expression (2) with parameters $\mu_{r,max} = 119$, $B_c = 0.4$ T and $\alpha = 1.3$. Alternatively to this expression, we have used the interpolation of experimental data (symbols) in the calculation of critical state distributions, without an observable difference for the calculated distributions. Another useful information that can be extracted from the magnetization data is the volume density of hysteresis loss in Ni layer determined by the areas of minor loops. The result of such analysis based on experimental data is plotted in the right hand side of Figure 10, together with the approximation

$$\boldsymbol{Q_{Ni}} = \begin{cases} \boldsymbol{Q_{sat}}\left(\frac{\boldsymbol{B}}{\boldsymbol{B_s}}\right)^{2} & \boldsymbol{B} \leq \boldsymbol{B_s} \\ \boldsymbol{Q_{sat}} & \boldsymbol{B} > \boldsymbol{B_s} \end{cases} \tag{26}$$

where $B_s = 0.5$ T and $Q_{sat} = 2750$ J/m$^3$.

With such complete description of the sample geometry, properties of superconductor and of ferromagnetic cover, it is possible to perform the numerical calculation of AC behaviour of our SC-FM composite wire. In fact all the distributions presented in the previous Section have been calculated for the actual geometry of the composite shown in Figure 9 and the material properties given in the preceding part of this Section. From the distributions calculated for the transport case the AC loss behaviour can be predicted and compared with experimental result, as shown in Figure 11. Empty circles indicate the transport loss measured on the original tape without ferromagnetic cover. The dashed line with square symbols is the result of AC loss calculation using the formula (25). The estimation for the loss in Ni cover on edges was carried out following the idea proposed by Grilli et al. [9]: Assuming that from the magnetic field distribution calculated for the maximum current $I_a$ the local magnetic field in ferromagnetic material will change monotonically until the state at $-I_a$, one can calculate local Ni loss during such process (that is one half of the loss in the whole AC cycle) utilizing the formula (26). Integrating the local loss over the whole volume of the Ni material gives the estimation of AC loss due to FM hysteresis shown by the dashed line with diamond symbols. Theoretical prediction for the total transport loss is given by summing the loss in superconductor with the Ni loss. Calculated result for our SC-FM composite, given by the solid line in Figure 11, is in qualitative agreement with the experimental data (triangles) particularly at currents approaching the critical current.

The same procedure was used to compare the theoretical prediction for the loss in our SC-FM tape exposed to AC field perpendicular to the long axis of the elliptic core. The result is shown in Figure 12, confirming reasonable validity of theoretical prediction. Interestingly, the cover of edges in the form presented here reduces significantly the AC loss due to superconductor hysteresis. In spite of the fact that the FM cover is an additional source of hysteresis loss and this loss prevails at low currents or weak AC fields, the total loss of the SC-FM composite is lower than that of the original tape. Because the loss due to Ni hysteresis is the main factor at low currents, there is further room for loss reduction when FM cover with narrower hysteresis loop would be used.

## 5. Conclusions

We have demonstrated both experimentally and by numerical calculation that adding of a Ni cover on the edges of a commercial Bi-2223/Ag tape reduces its AC loss in both transport and magnetization regimes.

Numerical procedure simulating the evolution of current and magnetic field distribution during the AC cycle has been developed for the purpose of theoretical prediction. Essential feature of this procedure is the use of the formulation proposed by A. Campbell [10] for the critical state in superconductor in terms of the vector potential of magnetic field. We have developed working formulas relating the current density with the properly calibrated vector potential in the manner compatible with finite element calculations in non-linear media. In this way, the calculated critical state distributions in superconductor take into account the modification of local magnetic fields by a non-linear ferromagnetic material.

With the help of the developed simulation procedure, two scopes not resolved yet can be reached: Theoretical prediction of AC loss in existing superconducting wires containing ferromagnetic parts – like YBCO coated conductors on ferromagnetic substrate or $MgB_2$ wires with Ni or Fe in the matrix - can be predicted. Also, optimization of the superconductor-ferromagnetic composite wires can be carried out in order to design the wires with improved properties.

Acknowledgement

This work was supported by the Slovak Research and Development Agency under the contract No. APVV-51-045605 and the European Commission under the contracts FU07-CT-2007-0051 and MRTN-CT-2006-035619.

Figure captions

Fig. 1: Schematic plot of the geometry used in our considerations. Superconducting wire with elliptic core (gray) defined by the semi-axes $a_{SC}$ and $b_{SC}$, respectively, is covered on edges by C-shaped ferromagnetic layer (black). Such composite is placed in the calculation box with radius $R$. Actually we used $R \approx 25 a_{SC}$.

Fig. 2: Comparison of the signum function (full line) with the hyperbolic tangent (dashed line).

Fig. 3: Current and field distributions calculated for the initial increase of the perpendicular magnetic field. From top to down the applied fields are 5 mT, 12 mT, 22 mT and 50 mT, respectively. Density of electrical current is characterized by the colour darkness, magnetic field is represented by the lines of constant vector potential.

Fig. 4: Series of distributions calculated for the magnetic fields reduced from the initial 12 mT (upper plot) through the intermediate values of 6 mT, 0 mT and -6 mT, respectively, down to -12 mT (lower plot).

Fig. 5: Magnetization loops evaluated from the series of current distributions.

Fig. 6: Distributions calculated for the values of transport current (from top to down) 18 A, 31 A, 51 A and 56 A, respectively.

Fig. 7: Series of distributions as calculated for the transport current reduced from the amplitude value of 50 A (the plot on top) to 25 A, 0 A, -25 A down to 50 A, respectively.

Fig. 8: Hysteresis loops of the magnetic flux surrounding the superconducting wire that transports electrical current. The part proportional to current has been subtracted from the data presented in the left plot to obtain the plot at the right side.

Fig. 9: Micrograph of the multifilamentary Bi-2223 wire with the edges covered by ferromagnetic layer of nickel (left) together with the simplified geometry used in the calculations.

Fig. 10: Properties of Ni layer - relative permeability and volume density of hysteresis loss – as determined from the magnetization data measured for 10 μm thick Ni layer of 1 mm Ag substrate at 77 K.

Fig. 11: Effect of Ni cover on the transport losses. Data show that the AC loss measured on the original uncovered tape (open circles) slightly reduced when the edges have been covered by Ni layer shown in Fig. 9 (full triangles). Our numerical method predicts that this is a combination of loss reduction in superconducting core (dashed line with open squares) with the additional dissipation due to Ni hysteresis (predicted by the dashed line with open diamonds). Theoretical prediction for the total loss of the composite, plotted by the full line, is in reasonable agreement with experiment.

Fig. 12: Effect of Ni cover on the magnetization losses. Data show that the AC loss measured on the original uncovered tape (open circles) significantly reduced when the edges have been covered by Ni layer shown in Fig. 9 (full triangles). Our numerical method shows a dramatic reduction of the loss in superconducting core (dashed line with open squares) – more than one order of magnitude at low fields. The additional dissipation due to Ni hysteresis (predicted by the dashed line with open diamonds) prevails at low fields and rises the theoretical prediction to the values plotted by the full line. Again the agreement between experiment and theoretical prediction is good.

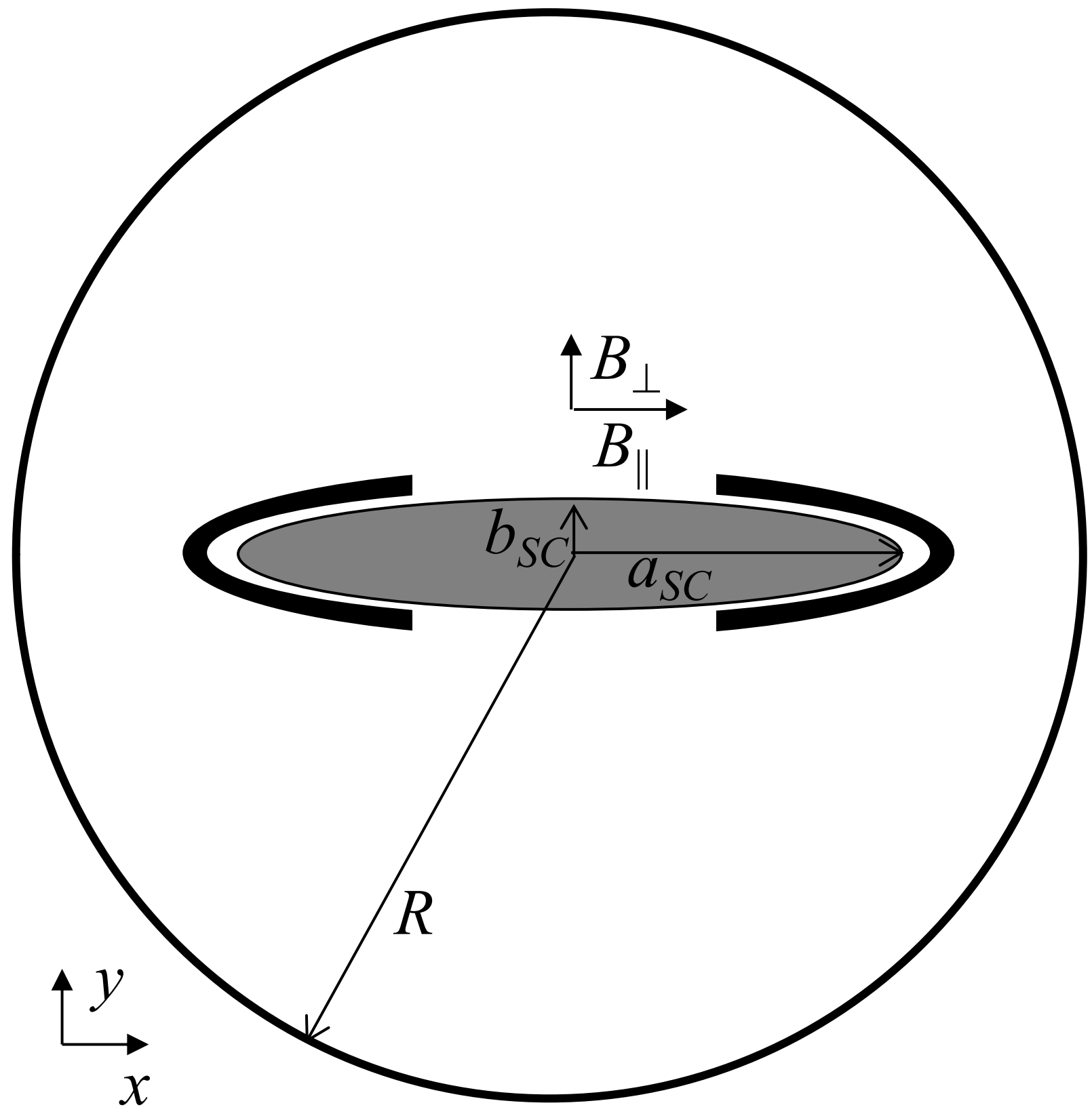

Figure 1

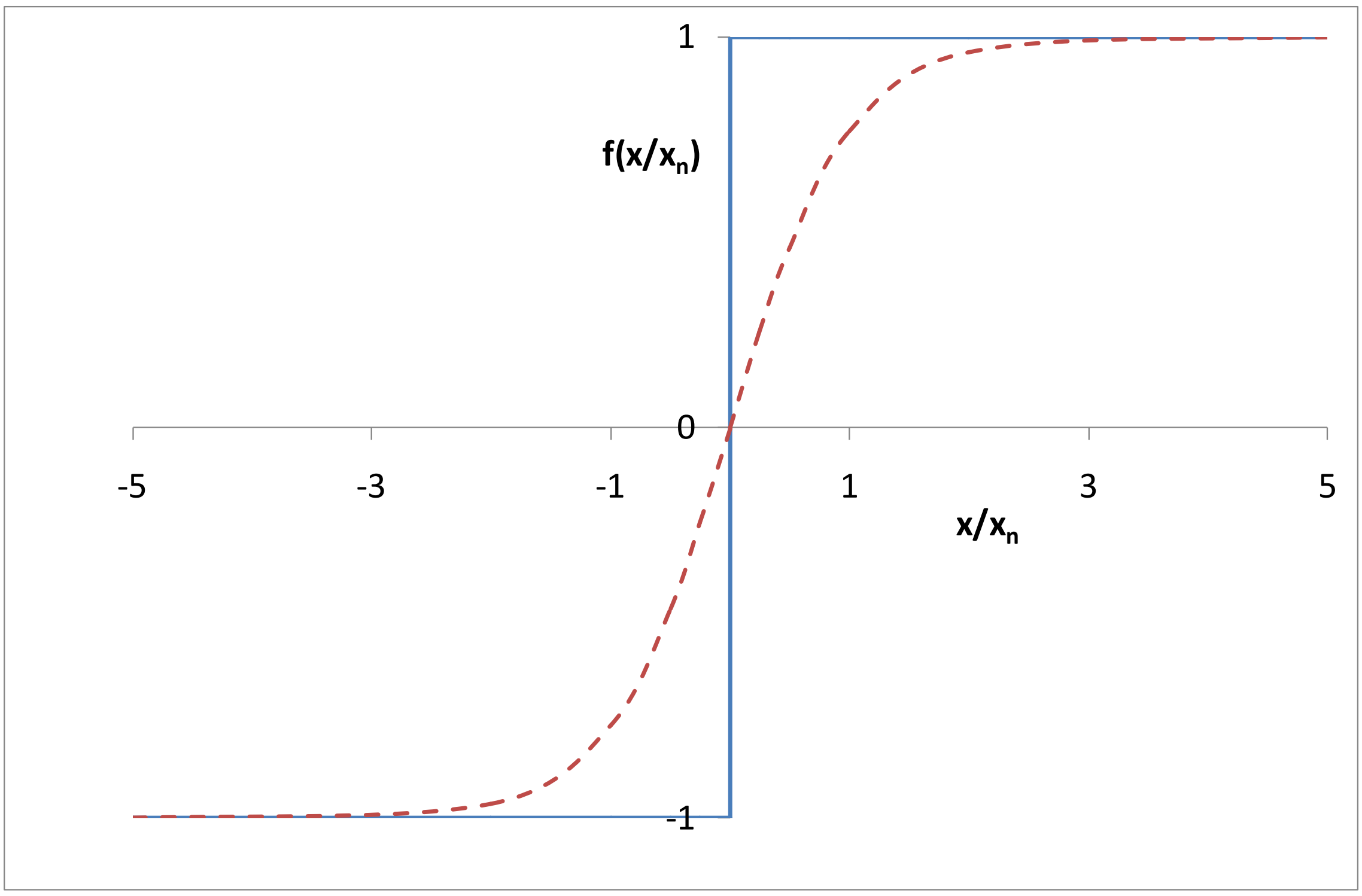

Figure 2

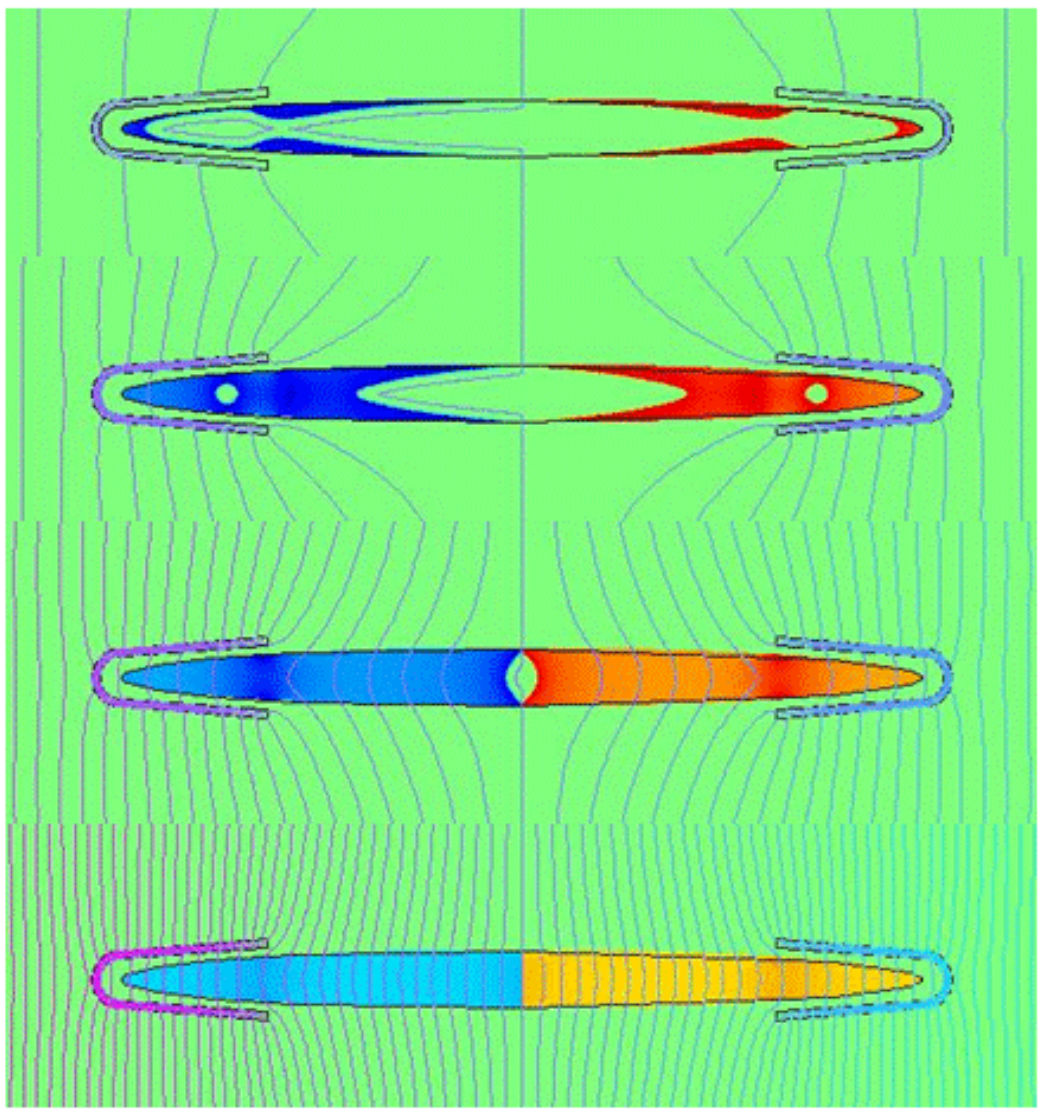

Figure 3

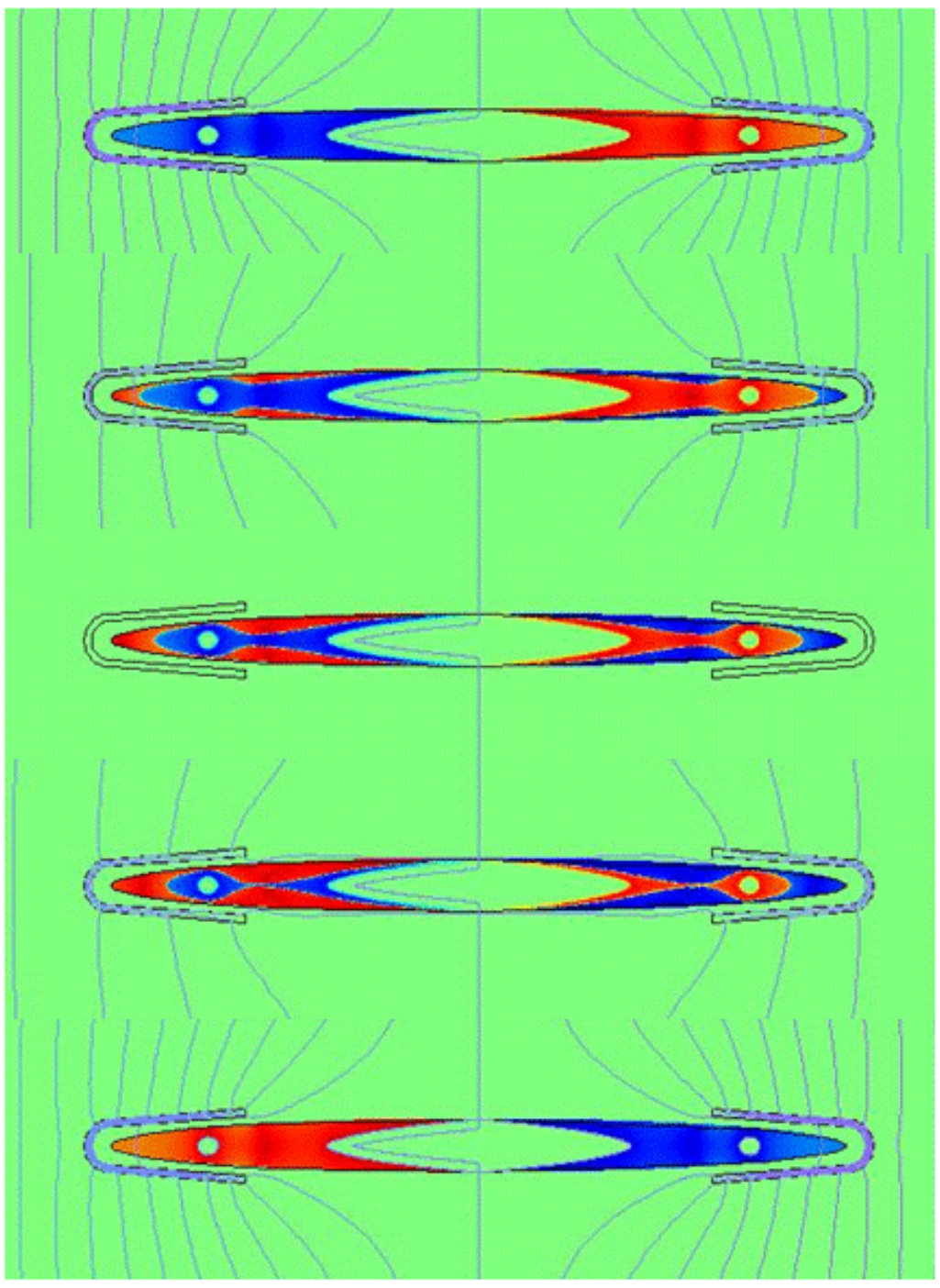

Figure 4

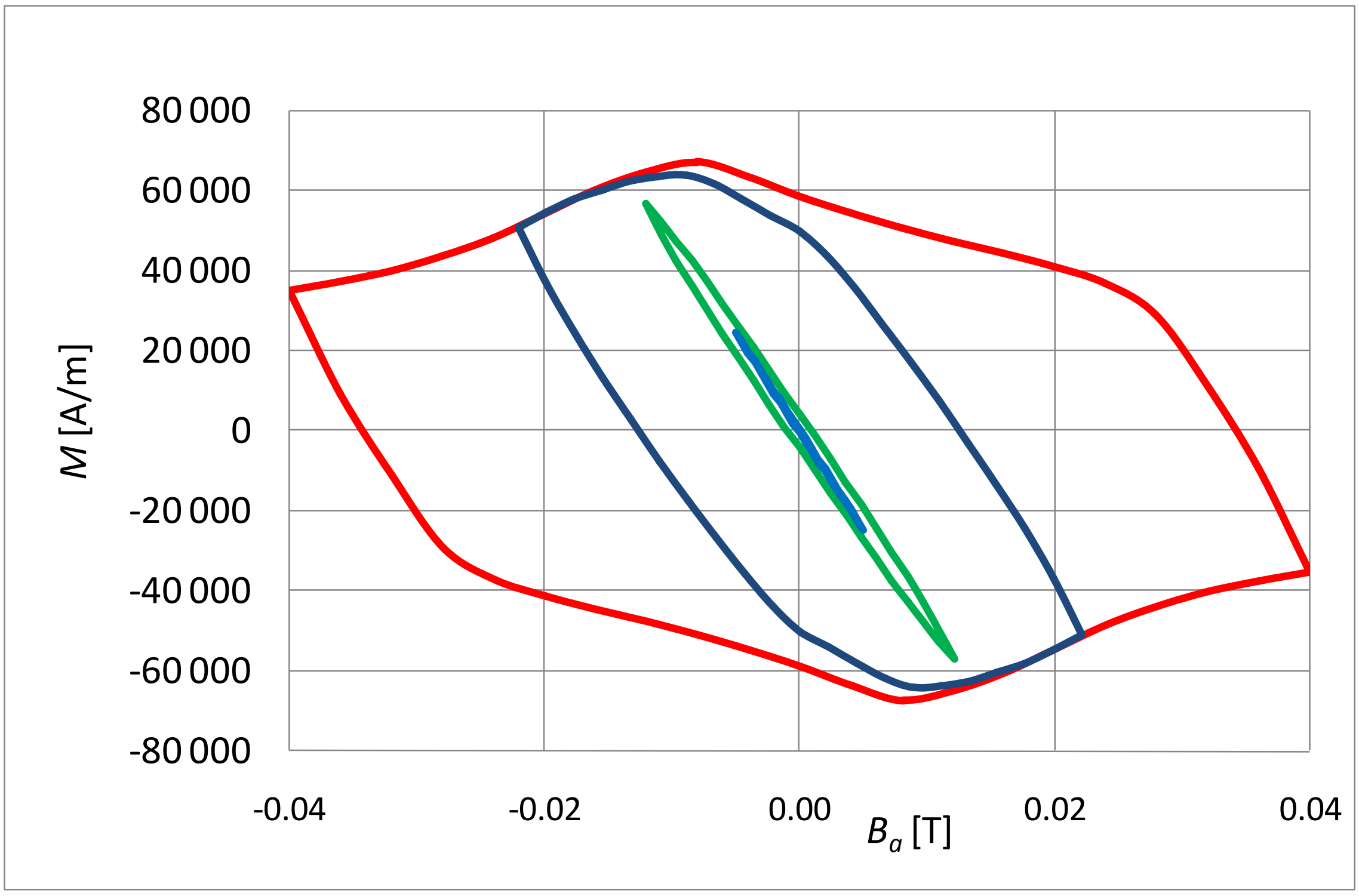

Figure 5

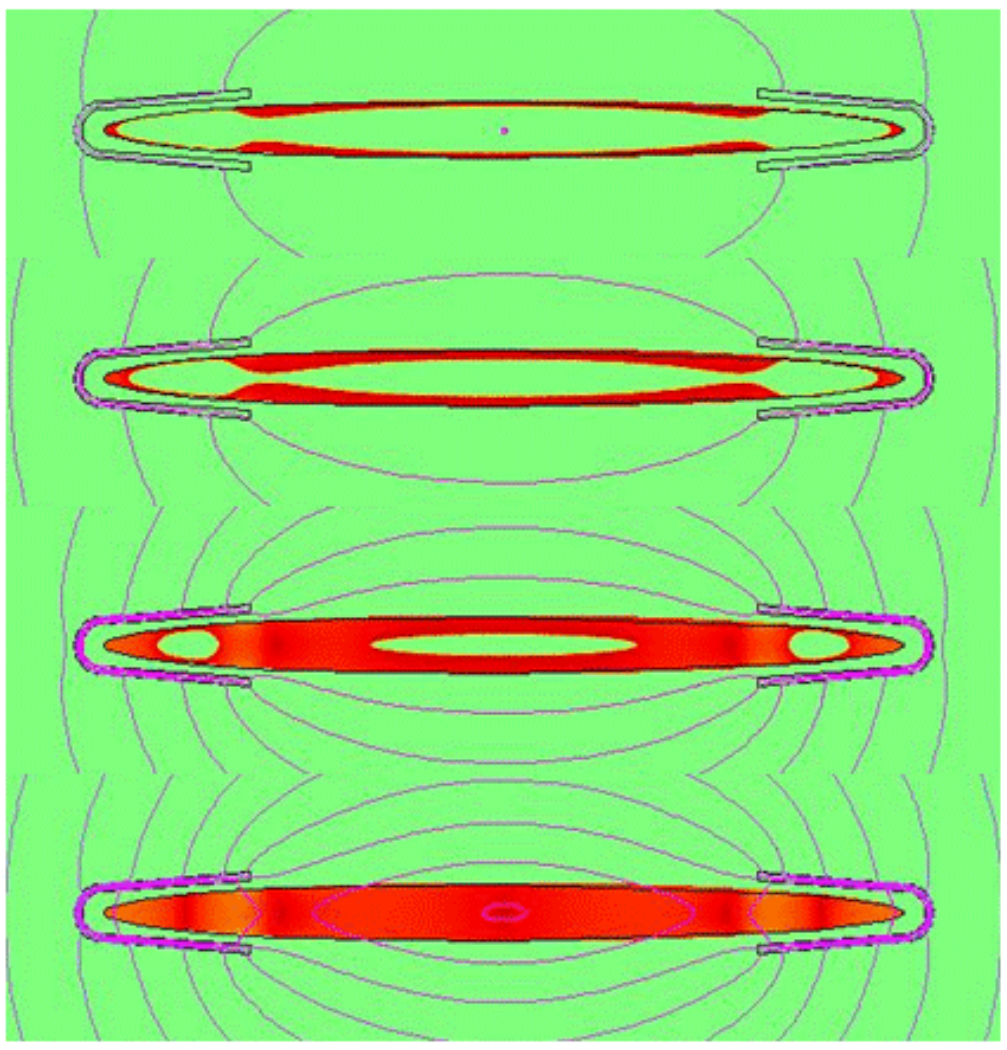

Figure 6

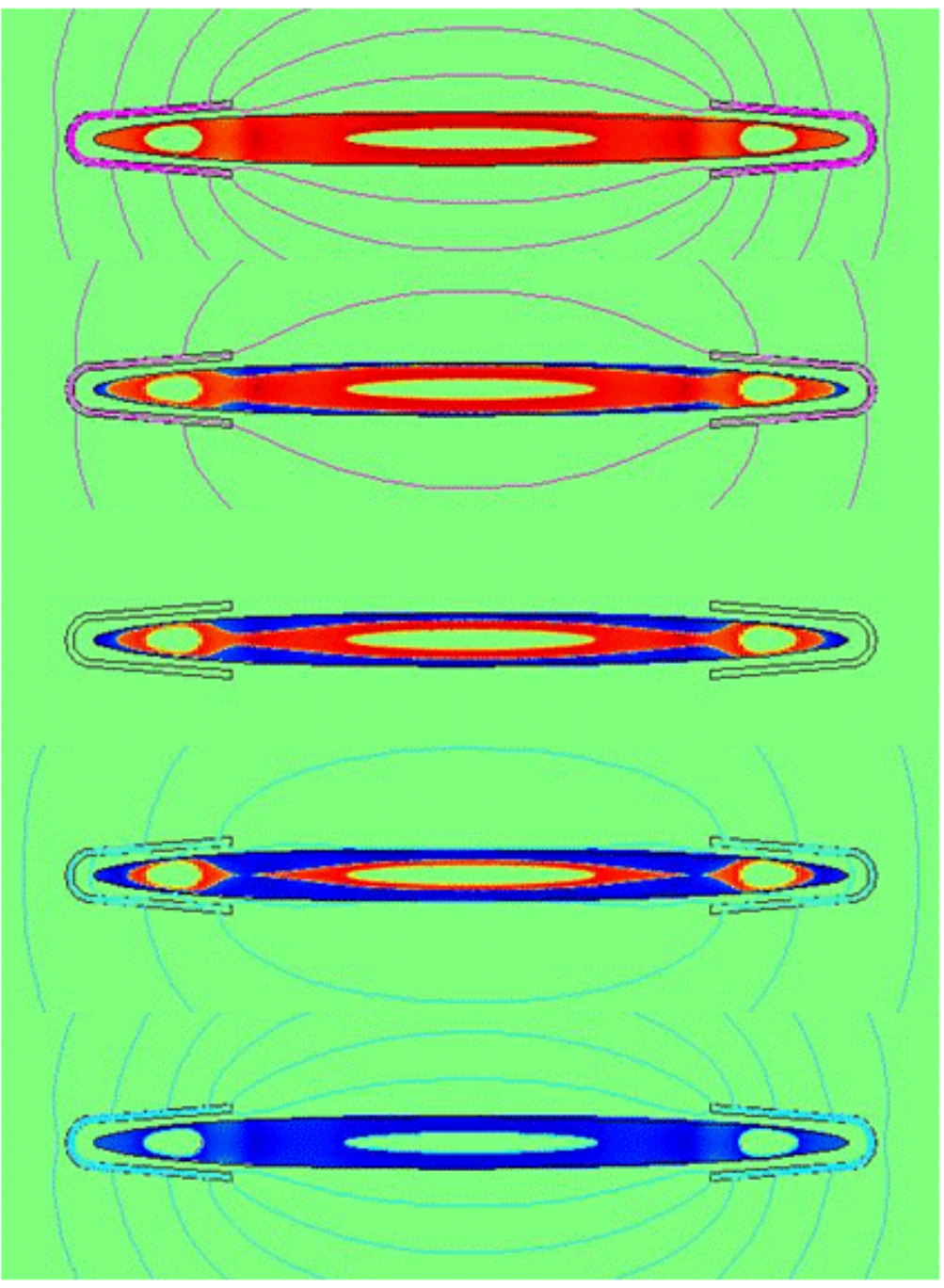

Figure 7

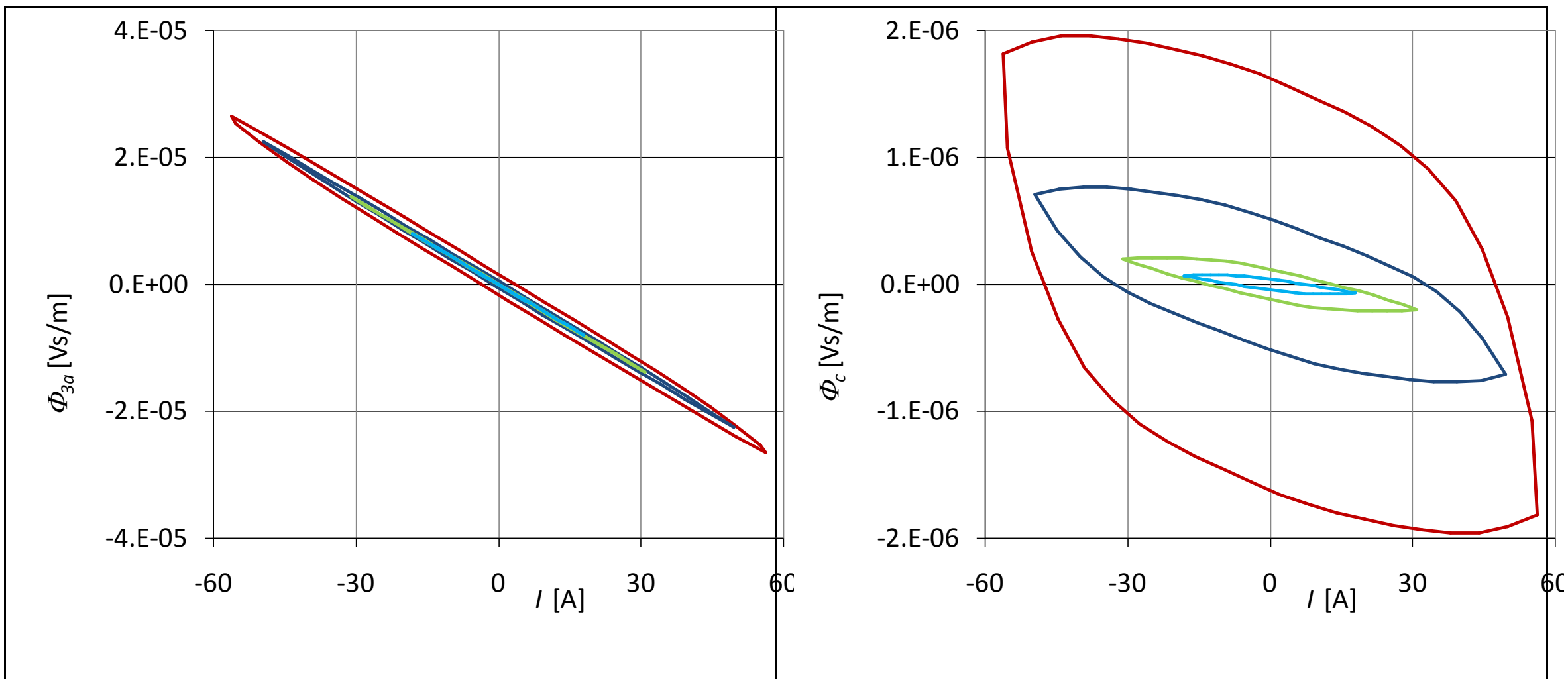

Figure 8

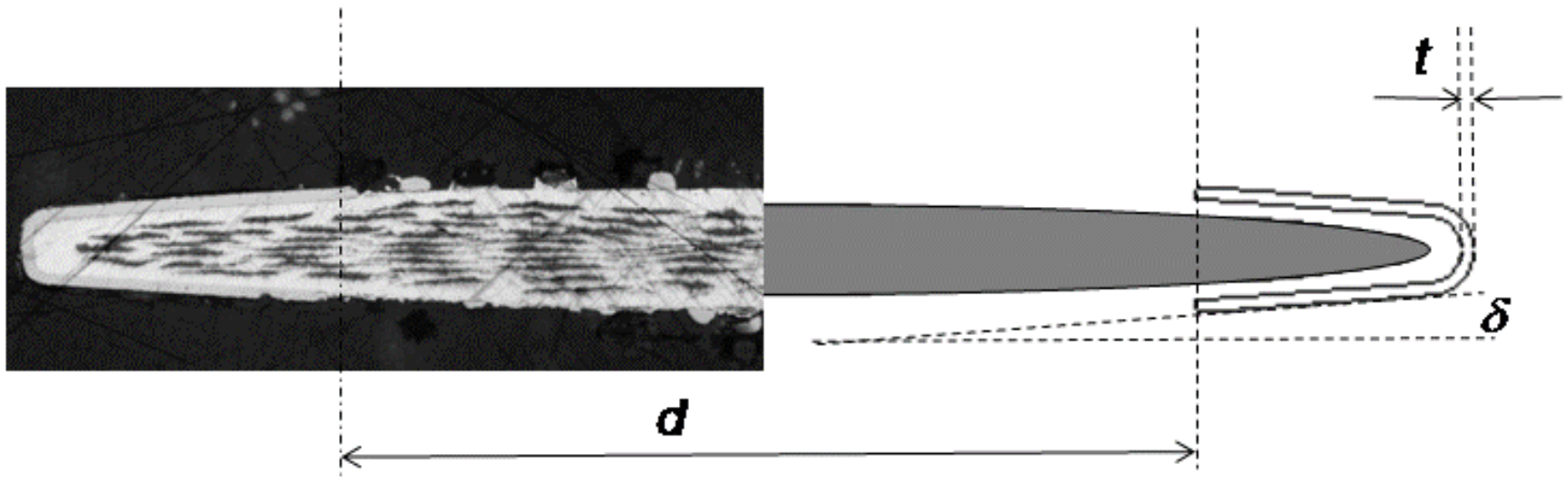

Figure 9

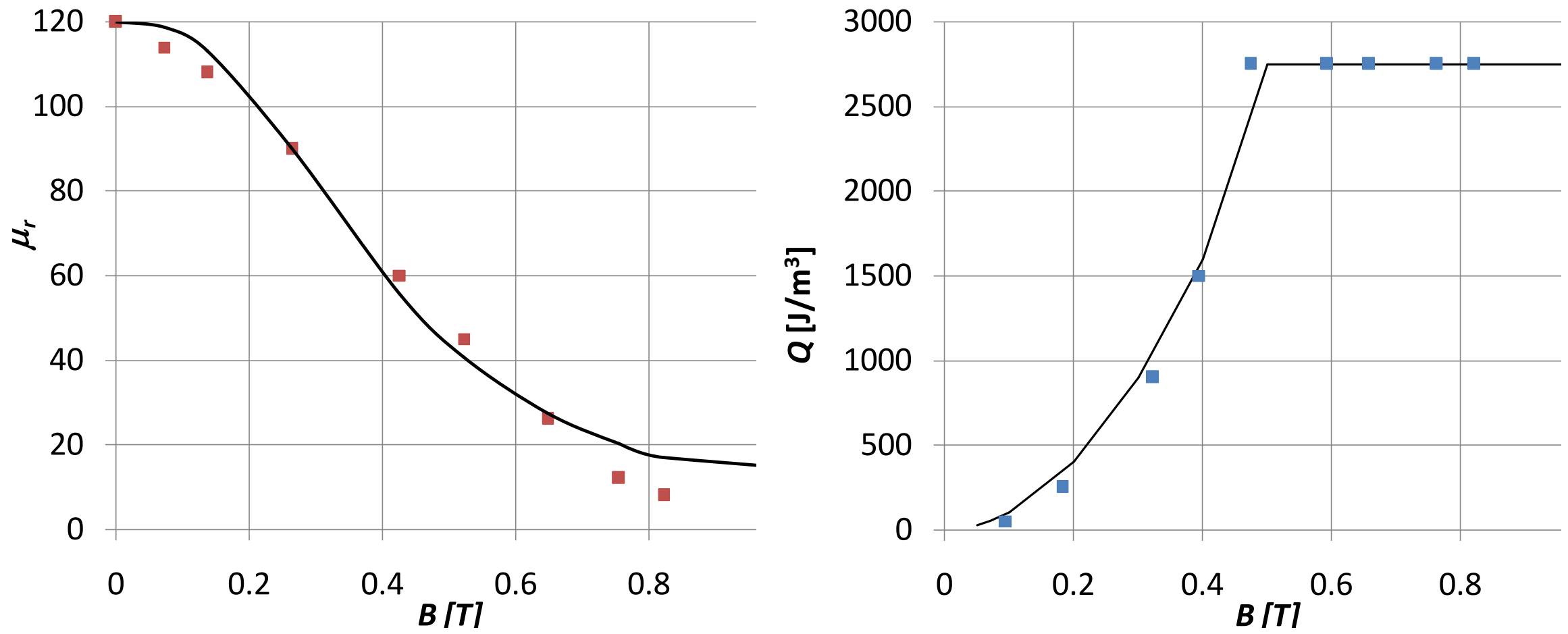

Figure 10

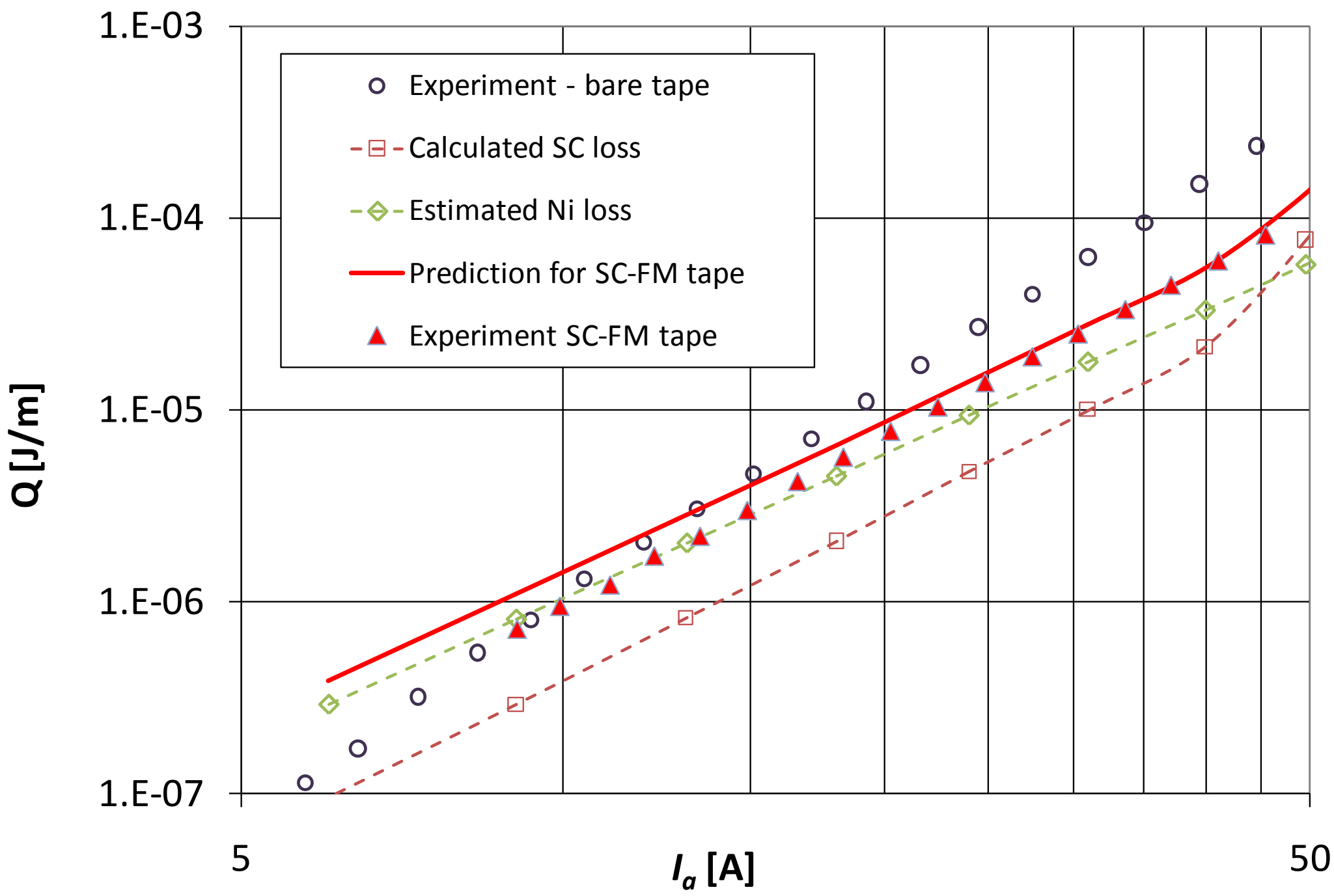

Figure 11

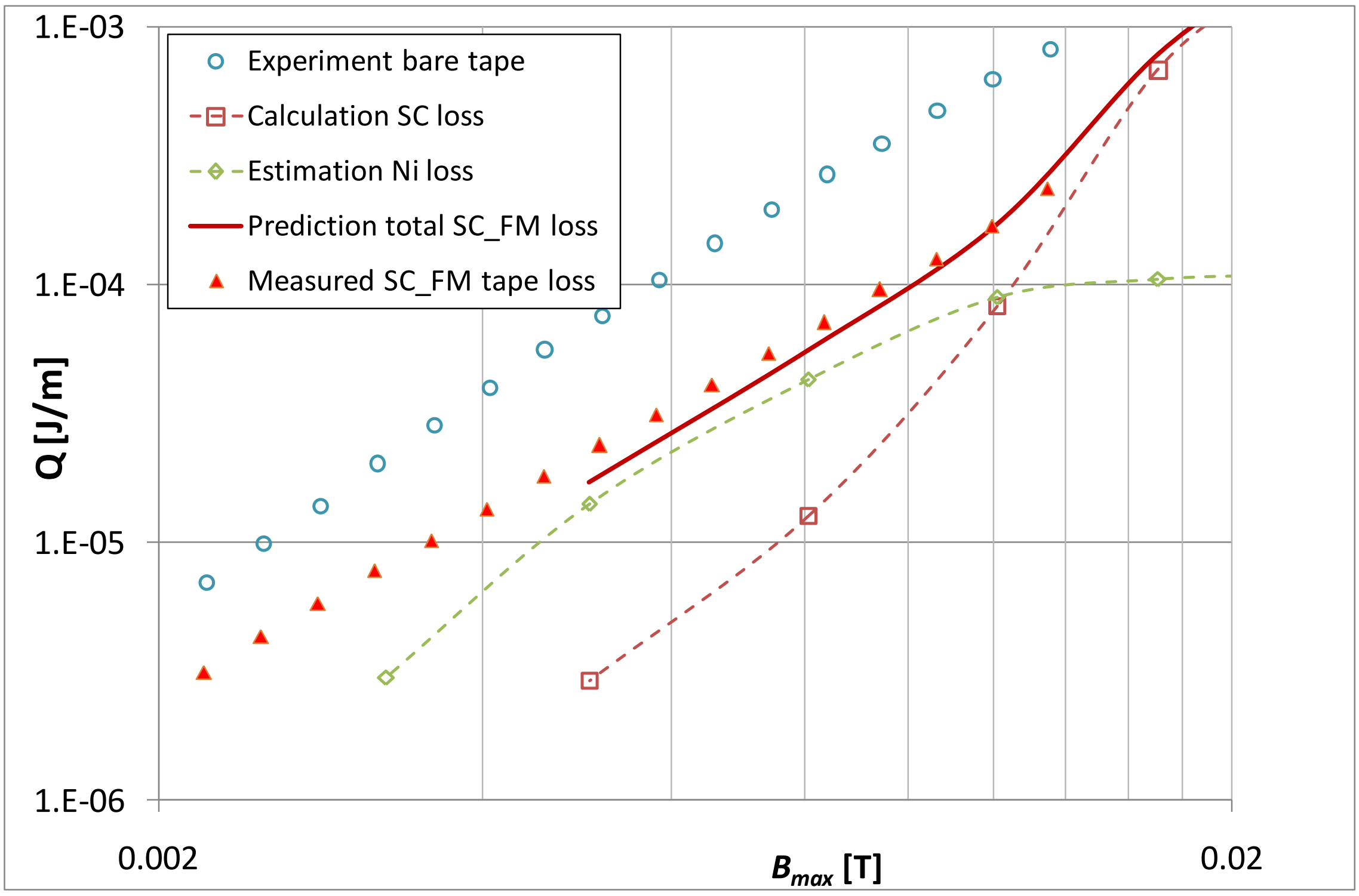
1.E-03
1.E-04
1.E-05
1.E-06
Q [J/m]
0.002
0.02
$B_{max}$ [T]
Experiment bare tape
Calculation SC loss
Estimation Ni loss
Prediction total SC_FM loss
Measured SC_FM tape loss

Figure 12